\documentclass[12pt,oneside, italian]{article}

\usepackage[english]{babel}

\usepackage{latexsym}

\usepackage{amssymb,amsthm,amsmath}

\usepackage{graphicx}

\title{\bf New quantum caps in $PG(4,4)$}

\author{Daniele Bartoli, Stefano Marcugini, Fernanda Pambianco}

\newtheorem{theorem}{Theorem}[section]

{\theoremstyle{definition}

\newtheorem*{definition*}{Definition}

\newtheorem{definition}[theorem]{Definition}}

\newtheorem*{proposition*}{Proposition}

\newtheorem*{corollary*}{Corollary}

\newtheorem*{lemma*}{Lemma}

\date{}

\begin{document}

\maketitle \vspace*{-15mm} \noindent

\ \\

\ \\

\begin{abstract}

\noindent Calderbank, Rains, Shor and Sloane (see \cite{Sloane}) showed that error-correction is possible in the context of quantum computations. Quantum stabilizer codes are a class of additive quaternary codes in binary projective spaces, which are self-orthogonal with respect to the symplectic form. A geometric description is given in \cite{Bierbra}, where also the notion of quantum cap is introduced. Quantum caps correspond to the special case of quantum stabilizer codes of distance $d=4$ when the code is linear over $GF(4)$. In the present paper we review the translation from quantum error-correction to symplectic geometry and study quantum codes in $PG(4,4)$. We already know that in $PG(4,4)$ there exist quantum caps of sizes $10,12,14-27,29,31,33,35$ (see \cite{Tonchev08}), of size $40$ (see \cite{40Cap}) and $41$ (see \cite{41Cap}). In this paper we presents quantum caps, complete and incomplete, of different sizes and in particular construct complete quantum caps with 36 and 38 points. Moreover we also show that there are only two examples of non equivalent quantum caps of size $10$ and five of size $12$; we prove by exhaustive search that no $11, 37, 39$-quantum caps exist. Besides we show that 20 is the minimum size of the complete caps in PG(4,4) (see  \cite{tesi}, \cite{OrdineMinimo} and \cite{RappTecn}) and that a 20-complete cap is quantic.

\end{abstract}

\section{Introduction}
In the second half of 20-th Century the new frontiers of modern physics led to
the introduction of new ideas in information theory. 
In particular quantum mechanics has given rise to the concept of quantum information.\\
Our aim is to determine the spectrum of quantum caps in $PG(4,4)$, which correspond to special linear quantum codes.\\ 
Quantum mechanics is based on the Heisenberg uncertainty principle, which is expressed by the formula
$$\Delta x \Delta y \geq \frac{h}{4\pi},$$
where $\Delta x$ is the error over the position of an elementary particle, $\Delta y$ is the error over the momentum and $h$ is the Planck constant.\\
The principle states that it is not possible to know at the same time and with absolute precision 
the speed and the position of an elementary particle, like the electron. \\
The fundamental unit of quantum information is the \emph{quantum bit} (qubit), 
which is like a two states physical system ($0$ and $1$) on which the superposition principle acts. This principle states that more than one state is present in the system at the same time. 
Physically a qubit is a two state quantum system, like the electron spin (up and down). \\
The idea of using quantum mechanical effects to perform computations was first introduced by Feyman in the $1980$s \cite{Feynman}, when he discovered that classical computers could not simulate all the aspects of quantum physics efficiently.\\
In $1982$ Wooters and Zurek \cite{nocloning} showed that there exists no quantum procedure to duplicate information contained in a qubit. In fact if it were possible, one could determine for example the polarization of a photon, by first producing a beam of identically polarized copies and then measuring the Stokes parameters. The linearity of quantum mechanics forbids such replication for all quantum systems.\\
In $1985$ Deutsch \cite{Deutsch} showed that it is possible to implement any function which is computable by classical computer using registers of entangled qubits and arrays of quantum gates.\\
In $1994$ Shor \cite{algoritmo_scomposizione} presented an algorithm which can factor an integer in polynomial time.\\
One of the most important problem in constructing quantum computer is decoherence. In the process of decoherence some qubits become entangled with the environment and this makes the state of the quantum computer \emph{collapse}. The conventional assumption was that once one qubit has decohered, the entire computation of the quantum computer is corrupted and the result of the computation will not be correct. In $1995$ Shor \cite{Shor} analyzed the problem of reducing the effects of decoherence for information stored in quantum memory, using the quantum analog of error correcting codes, and presented a procedure to encode a single qubit in nine qubits which can restore the original state if no more than one qubit of a nine-tuple decoheres. It is an example of a quantum $[[9,1,3]]$-code.\\ 
In $1998$ Calderbank, Rains, Shor and Sloane  \cite{Sloane} translated the problem of finding quantum error correcting codes into the problem of finding additive codes over $GF(4)$ which are self-orthogonal with respect to a particular trace inner product.\\
A \emph{quantum code} in this context is a set of configurations of a certain number of qubits.\\
This new type of codes has only recently made its appearance in coding theory: 
a classical code $\mathcal{C}$ is determined
by three parameters $n,k,d$ which measure length, dimension (i.e. the number of codewords)
and minimum distance of the code (which gives a rating of the number of errors the code can correct) respectively.\\
The main problem of coding theory is the optimization of one of these parameters 
when the others are fixed; for example maximizing the minimum distance 
for a fixed length and dimension. \\
It is possible (see \S \ref{sez:fisica}) to translate the description of quantum codes in terms of configurations of qubits to a description in terms of points in finite projective spaces. 
In the projective space $PG(r,q)$ over the Galois Field $GF(q)$, a $n$-cap is a set of $n$ points no $3$ of which are collinear. A $n$-cap is called \emph{complete} if it is not contained in a $(n+1)$-cap. \\
We call an $n$-cap a $n$-\emph{quantum} cap if the code generated by its matrix is a 
\emph{quantum stabilizer code} (see Definitions \ref{def:quantumcodes} and \ref{parametri}).\\
In $1999$ Bierbrauer and Edel showed that $41$ is the maximum size of complete caps in $PG(4,4)$  and this cap is quantic (see \cite{41Cap}). In $2003$ the same authors presented a complete $40$-cap in $AG(4,4)$ which is also quantic (see \cite{40Cap}).\\ 
In $2008$ Tonchev constructed quantum caps of sizes $10,12,14-27,29,31,33,35$ (see \cite{Tonchev08}), starting from the complete $41$-quantum cap in $PG(4,4)$ (see \cite{41Cap}).\\
It is not difficult to see (\cite{AMS}) that this method cannot produce quantum caps of sizes between $36$ and $40$ in $PG(4,4)$.\\
In $2009$ we have found examples of quantum caps of sizes $13,28,30,32,34,36,38$, see \cite{AMS}.\\
Using the theoretical results of Section \ref {sez:fisica}, we determine in Section \ref{spectrum}, by a computer based search,  the spectrum of size of quantum caps in $PG(4,4)$ (and therefore of pure linear quantum $[[n,n-10,4]]$-codes) proving that there exist no examples of quantum caps of sizes $11$, $37$ and $39$. Then we proved the following:
\begin{theorem}
If $\mathcal{K} \subset PG(4,4)$ is a quantum cap, then $10\leq |\mathcal{K}| \leq 41$, with $|\mathcal{K}| \neq 11,37,39$.
\end{theorem}
With the same means we prove that the minimum size of complete caps in $PG(4,4)$ is $20$. In the search for quantum caps we have utilized theoretical results (see Section \ref {sez:fisica}) and a searching algorithm (see Section \ref{spectrum}).\\
In Section \ref{sez:fisica} we start from the physical description of quantum codes in order to better understand their mathematical definition in terms of classical coding theory. 
In Section \ref{sez:teo} we give some theoretical results which have been utilized in the computer-based research of quantum caps in $PG(4,4)$ (see \S \ref{alg}).

In Section  \ref{risultati} we present the spectrum of linear pure quantum codes of type $[[n,n-10,4]]$ and determine the minimum size of complete caps in $PG(4,4)$.

\section{From physical to mathematical description of quantum codes}\label{sez:fisica}

In the context of quantum physics  a \emph{quantum code} is a set of
configurations of a certain number of qubits. All the physical features of quantum codes can be translated into a mathematical
setting. For this purpose a qubit can be considered as an element of a
two dimensional complex Hilbert space $\mathcal{H}$. The two qubit
base-states (\emph{kets}) are 
$$|0\rangle=\left( \begin{array}{c} 0\\
1 \end{array} \right)  \textrm{ and } |1\rangle=\left( \begin{array}{c} 1\\ 0
\end{array} \right).$$
A general qubit is a linear combination of the
two base-states, as $\alpha |0\rangle + \beta |1\rangle,$ where  $\alpha$ and $\beta$ are complex numbers such that $|\alpha|^{2}+|\beta|^{2}=1$. 
Previous notation represents the superposition principle of the base-states $|0\rangle $ and $|1\rangle$. \\
In general we can consider a system of $n$ qubits as an element of the $n$ times tensorial product of 
$\mathcal{H}$. A quantum code $\mathcal{C}$ is determined by a set of particular base-configurations of some qubits 
(for example $n$). Since qubits behave totally differently from classical bits, Nielsen and Chuang \cite{3problemi} summarize three difficulties in quantum error correction (see also \cite{intro2}):

\begin{enumerate}

\item measurement destroys informations: in fact it is not possible to know the phases $\alpha$ and 
$\beta$ of a single qubit. If we do a  measurement, we obtain 0 with probability $|\alpha|^{2}$ and 1 
with probability $|\beta|^{2}$;
\item \emph{No cloning theorem} states that no quantic procedure to duplicate informations exists;
\item qubit errors are a \emph{continuum}.
\end{enumerate}
To solve the first problem some techniques similar to the syndromes in classic coding theory are utilized; 
to solve the second problem we embed the configurations of qubits in Hilbert space of 
larger dimension; to solve the third problem we consider errors as operators in Hilbert spaces.\\
In particular one way to solve the problem that qubit errors are a continuum is to view a single error like an operator in a Hilbert space $\mathcal{H}$, 
i.e. a $(2 \times 2)$-complex matrix. As a quantum code is a set of particular base-configurations of 
$n$ qubits, we can view an error like the tensorial product of $n$ operators, each of them acting on a 
single qubit. Each single operator is the linear combination of the \emph{Pauli matrices}:

\begin{displaymath}
\mathbb{I}= \left (
\begin{array}{cc}
1&0\\
0&1\\
\end{array}
\right)
\sigma_{x}= \left (
\begin{array}{cc}
0&1\\
1&0\\
\end{array}
\right)
\sigma_{y}=\left(
\begin{array}{cc}
0&-i\\
i&0\\
\end{array}
\right)
\sigma_{z}= \left (
\begin{array}{cc}
1&0\\
0&-1\\
\end{array}
\right).
\end{displaymath}
These matrices act on a single qubit in the following way;

\begin{center}

\begin{tabular}{|c|c|c|}

\hline

Identity&$\mathbb{I}$& $\mathbb{I}|a\rangle=|a\rangle$\\

\hline

Bit Flip&$\sigma_{x}$&$\sigma_{x}|a\rangle =|a\oplus 1 \rangle$\\

\hline

Phase Flip&$\sigma_{z}$&$\sigma_{z}|a\rangle =(-1)^{a}|a \rangle$\\

\hline

Bit and Phase Flip &$\sigma_{y}$&$\sigma_{y}|a\rangle =i(-1)^{a}|a\oplus 1 \rangle$\\

\hline

\end{tabular}

\end{center}
Then we can give now a description of errors using a finite set of base-errors and not a 
\emph{continuum}. In the following if $E$ is an error and $\psi$ a base-codeword, $E|\psi$ represents the error-operator $E$ acting on $\psi$. We can demonstrate that the set of all the error-operators is a vectorial space 
and that a quantum code with base-codewords $\psi_{i}$ and errors $E_{a}$ must satisfy the equations
$$\langle \psi_{i}|E_{a}^{H},E_{b}|\psi_{j}\rangle=0 \quad \forall i\neq j \qquad \textrm{ and } \qquad \langle \psi_{i}|E_{a}^{H},E_{b}|\psi_{i}\rangle=\langle 
\psi_{j}|E_{a}^{H},E_{b}|\psi_{j}\rangle \quad \forall i, j,$$ where $\langle , \rangle$ is the inner product in the considered Hilbert space and $E^{H}$ is the Hermitian matrix of $E$. It can also be proved (see \cite{Bennett} and \cite{intro1}) that the above equations are equivalent to
\begin{equation}\label{equazioni}
\langle \psi_{i}|E_{a}^{H},E_{b}|\psi_{j}\rangle=C_{ab}\delta_{ij}\quad \forall i, j,
\end{equation}
where $\psi_{i}$ and $\psi_{j}$ are all the possible base-codewords, $E_{a}$ and $E_{b}$ are errors, $C_{ab}$ does not depend on $i$ and $j$ and $\delta$ is the Kronecker symbol. These conditions are also sufficient for the existence of a 
code and a set of errors which respect the uncertainty principle.\\
The most utilized quantum codes are the \emph{quantum stabilizer codes}. 
Let $\mathcal{C}$ be a set of possible quantic configurations of $n$ qubits. 
Let $\mathcal{G}$ be the set of all error-operators and let
\begin{displaymath}
\mathcal{S}=\{ E \in \mathcal{G} \textrm{ } |\textrm{ } E|\psi\rangle=|\psi\rangle \textrm{ } 
\forall \psi \in \mathcal{C}\},
\end{displaymath}
be the set of the operators which fix all the codewords. In particular all the codewords are 
eigenvectors of each error-operator whit eigenvalue 1. In general, the stabilizer $\mathcal{S}$ is an abelian subgroup of $\mathcal{G}$ and the code 
$\mathcal{C}$ is the space of the vectors fixed by $\mathcal{S}$.\\
If we take two generic Pauli matrices, they can only commute or anticommute and therefore it is easy 
to see that each error-operator, which is the tensorial product of Pauli matrices, can only commute 
or anticommute too. \\
Let $E$ be an error-operator which anticommutes with a certain $M \in \mathcal{S}$ (i.e. $M E+E M=0$). 
Then we have  $ME|\psi_{i}\rangle=-EM|\psi_{i}\rangle=-E|\psi_{i}\rangle$, as $M$ fixes every codeword. 
As $E|\psi_{i}\rangle$ is an eigenvector with eigenvalue -1 of $M$, then $E|\psi_{i}\rangle$ 
is not a codeword and there has been an error. Otherwise if $E$ commutes with any element $M \in \mathcal{S}$, then 
$ME|\psi_{i}\rangle=EM|\psi_{i}\rangle=E|\psi_{i}\rangle$ and it is not possible to know if 
any error occurred, as $E|\psi_{i}\rangle$ is an eigenvector of $M$ with eigenvalue 1 
(like the other codewords). It is possible to demonstrate (see \cite{Bennett} and \cite{intro1}) that a set of base-state configurations of a certain number of qubits 
$\mathcal{C}$ and a set of error-operators $\mathcal{E}$ such that each  $E=E_{a}^{H}E_{b}$ 
anticommutes with some  $M \in \mathcal{S}$, with $E_{a}, E_{b} \in \mathcal{E}$, verify the equations 
\ref{equazioni} and then $\mathcal{C}$ defines a quantum code which corrects each error of 
$\mathcal{E}$. However it is not possible to correct the error-operators which commute with all the elements of 
$\mathcal{S}$. Let $C(\mathcal{S})$ be the set of all the operators which commute with the elements 
of $\mathcal{S}$ and $N(\mathcal{S})=\{ \omega \textrm{ }| \textrm{ } \omega \mathcal{S} \omega^{-1}= \mathcal{S}\}$. We can demonstrate that the two sets are the same 
and the stabilizer quantum code can correct all the errors of the set $\mathcal{E}$, such that 
$E_{a}^{H}E_{b} \in \mathcal{S} \cup (\mathcal{G} \setminus N(\mathcal{S})) \textrm{ } \forall E_{a},E_{b} \in \mathcal{E}$.\\
If the quantum code $\mathcal{C}$ encodes $k$ qubits to $n$ qubits, then a set of generators for its 
stabilizer has size $n-k$. We can also translate each Pauli matrix to an element of $GF(2)^{2}$:
$\sigma_{x}  \to 10$, $\sigma_{y} \to  11$, $\sigma_{z}  \to  01$ and $\mathbb{I}  \to  00$.
This translation has the property that two Pauli matrices commute
iff the symplectic product (i.e. $f((x_{1},y_{1}),(x_{2},y_{2}))=x_{1}y_{2}+x_{2}y_{1} \in GF(2)$) of their translations is equal to 0, and
anticommute if and only if the symplectic product is equal to 1. A generic
operator in $\mathcal{S}$ is the tensorial product of Pauli matrices
and the product of two operators is:
$$(A_{1} \otimes \ldots \otimes A_{n})\times (B_{1} \otimes \ldots \otimes B_{n})=(A_{1}\times B_{1}) 
\otimes \ldots \otimes (A_{n}\times B_{n}),
$$
by the properties of tensorial product, where $A_{i}$ and $B_{j}$ are
the matrices corresponding to base-errors $\sigma_{x}, \sigma_{y},\sigma_{z}, \mathbb{I}$. The product is then
\begin{displaymath}
(A_{1}\times B_{1}) \otimes \ldots \otimes (A_{n}\times B_{n})=(-1)^{k}(B_{1}\times A_{1}) 
\otimes \ldots \otimes (B_{n}\times A_{n}),
\end{displaymath}
where $k$ is the number of times that a single base-error anticommutes. 
Two operators commute iff the number of indices corresponding to Pauli matrices which anticommute 
is even and viceversa they anticommute iff this number is odd. \\
If we consider  $\mathbf{F}=GF(2)$ and $\mathbf{V}=\mathbf{F}^{2n}$, we can represent each element 
$\omega$ of $\mathbf{V}$ like
\begin{displaymath}
\omega=(x_{1}y_{1},x_{2}y_{2},\ldots,x_{n}y_{n}) \quad \textrm{with} \quad x_{i},y_{i} \in \mathbf{F} 
\quad \forall i=1,\ldots,n.\end{displaymath}
In general we define \emph{symplectic form} the function $\Phi : \mathbf{V} \times \mathbf{V} \to \mathbf{F}$ defined by
\begin{displaymath}
\Phi(\omega_{1},\omega_{2})  =  \Phi ((x_{1,1}y_{1,1},x_{1,2}y_{1,2},
\ldots,x_{1,n}y_{1,n}),(x_{2,1}y_{2,1},x_{2,2}y_{2,2},\ldots,x_{2,n}y_{2,n}))=
\end{displaymath}
\begin{displaymath}
\sum_{i=1}^{n}(x_{1,i}y_{2,i}-y_{1,i}x_{2,i})
\end{displaymath}
The symplectic form satisfies:
$$f(\alpha_{1} x_{1}+ \alpha_{2} x_{2},y) =\alpha_{1}f(x_{1},y)+\alpha_{2}f(x_{2},y),\quad  f(y,x)=-f(x,y),\quad  f(x,x)=0.$$
We can see that two error-operators commute or anticommute if the symplectic product of the respective 
translations is equal to 0 or 1.\\
Then we can build a matrix whose rows are obtained by translating
the generators of the stabilizer
 of a quantum code $\mathcal{C}$. If this code encodes $k$ qubits in $n$ qubits,
 the matrix is a $(n-k , 2n)$-binary matrix and the rows are orthogonal to each other with
 respect to the symplectic form. There exist some elements which are not linear combination
 of matrix's rows (i.e. they are not elements of $\mathcal{S}$) that commute with every rows: 
they are elements of $N(\mathcal{S}) \setminus \mathcal{S}$. 
As we have seen above they are not correctable by the code. We can indicate $N(\mathcal{S})$ with 
$\mathcal{C}^{\bot}$, because in this set there are all the elements which are orthogonal with respect 
to the symplectic form to all the elements of $\mathcal{S}$.\\

\section{Theoretical background}\label{sez:teo}

A linear $q$-ary $[n,k]$-code $\mathcal{C}$ is a $k$-dimensional subspace of $GF(q)^{n}$. 
A $q$-linear  $q^{m}$-ary $[n,k]$-code is a $km$-dimensional $GF(q)$-subspace of $GF(q)^{mn}$. 
In particular an \emph{additive} code $\mathcal{C}$ over $GF(4)$ is a subset of $GF(4)^{n}$ 
closed under addition.
By the above considerations this definition follows (see \cite{tesi} and \cite{Sloane}):
\begin{definition}\label{def:quantumcodes}
A \emph{quaternary quantum  stabilizer code} is an additive quaternary code $\mathcal{C}$ 
contained in its dual $\mathcal{C}^{\bot}$, where the duality is with respect to the symplectic form (see \S $\textrm{ }$ \ref{sez:fisica}).
\end{definition}
\noindent In particular:
\begin{definition}\label{parametri}
A quantum code  $\mathcal{C}$ with parameters $n,k,d$ ($[[n,k,d]]$-code), where $k>0$, 
is a quaternary quantum stabilizer code of binary dimension $n-k$ satisfying the following: 
any codeword of $\mathcal{C}^{\bot}$ having weight at most $d-1$ is in $\mathcal{C}$.\\
The code is \emph{pure} if $\mathcal{C}^{\bot}$ does not contain codewords of weight $< d$, 
equivalently if $\mathcal{C}^{\bot}$ has strength $t \geq d-1$.\\
An $[[n,0,d]]$-code $\mathcal{C}$ is a self-dual quaternary quantum stabilizer code of strength 
$t=d-1$.
\end{definition}
If we describe an $[[n,k,d]]$-code $\mathcal{C}$ by a generator matrix, we can consider each of the 
$n$ coordinate sections containing 2 columns. A generator matrix is for example the following:\\
\begin{displaymath}
\left ( \begin{array}{cccc}
P_{1,1}Q_{1,1} & P_{1,2}Q_{1,2} & \ldots&P_{1,n}Q_{1,n}\\
P_{2,1}Q_{2,1} & P_{2,2}Q_{2,2} & \ldots &P_{2,n}Q_{2,n}\\
\vdots& \vdots & &\vdots\\
P_{n-k,1}Q_{n-k,1} & P_{n-k,2}Q_{n-k,2} & \ldots &P_{n-k,n}Q_{n-k,n}\\
\end{array} \right)
\end{displaymath}
with $P_{i,j},Q_{i,j} \in \mathbb{Z}_{2}$ $\forall i=1,\ldots,n-k \quad j=1,\ldots,n$. \\
We can view each column as a point in the binary projective space $PG(n-k-1,2)$. 
Hence the geometric description of the quantum code is in terms of a system of $n$ lines 
(\emph{codelines}) generated by the $n$ pairs of points. However it is possible that the 
0-column occurs and that two different columns in the same coordinate section are identical. \\
For a more detailed introduction to quantum codes see in particular \cite{Bierbra}, \cite{intro1} 
and \cite{intro2}.\\
The following theorem gives a first geometrical description of quantum codes.
\begin{theorem}\label{th:caratt}
The following are equivalent:
\begin{itemize}
\item a \emph{pure} quantum $[[n,n-m,t+1]]_{2}$-code;
\item a set of $n$ lines in $PG(m-1,2)$ any $t$ of which are in general position 
and such that for each $\mathbf{secundum}$ $S$ (subspace of codimension 2) the number of lines which are skew to $S$ is even.
\end{itemize}
\end{theorem}
\proof
Let $\mathit{x}$ and $\mathit{y}$ two codewords, $\mathit{x}=A_{h_{1}}+\ldots+A_{h_{j_{1}}}$ and $\mathit{y}=A_{i_{1}}+\ldots+A_{i_{j_{2}}}$,  where $A_{h}$ is the $h$-th row of the generator matrix. We can associate to them two hyperplanes 
$\textit{H}_{1}$ and $\textit{H}_{2}$ in $PG(m-1,2)$ with equations
\begin{displaymath}
H_{1} : x_{h_{1}}+\ldots+x_{h_{j_{1}}}=0 \qquad  \textrm{ and } \qquad H_{2} : x_{i_{1}}+\ldots+x_{i_{j_{2}}}=0.
\end{displaymath}
We can suppose that the $k$-th entries of  $\mathit{x}$ and $\mathit{y}$ are the pairs
\begin{displaymath}
(P_{h_{1},k}+\ldots+P_{h_{j_{1}},k},Q_{h_{1},k}+\ldots+Q_{h_{j_{1}},k})=(H_{1}(P_{k}),H_{1}(Q_{k}))
\end{displaymath}
and
\begin{displaymath}
(P_{i_{1},k}+\ldots+P_{i_{j_{2}},k},Q_{i_{1},k}+\ldots+Q_{i_{j_{2}},k})=(H_{2}(P_{k}),H_{2}(Q_{k})).
\end{displaymath}
Let $S$ be the secundum $\textit{H}_{1} \cap \textit{H}_{2}$. \\
Let $L_{k}\nsubseteq S$ be a line such that $S \cap L_{k} \neq \emptyset$ and let $R_{k} \in S \cap L_{k} $. We can have only one
of the following situations:
\begin{itemize}
\item $R_{k}$ is $P_{k}$ or $Q_{k}$. Then the $k$-th entries of $\mathit{x}$ and $\mathit{y}$ are 
$(0,\alpha)$ and $(0,\beta)$ or $(\alpha,0)$ and $(\beta,0)$ with $\alpha, \beta \in \mathbb{Z}_{2}$, 
because $H_{1}(R_{k})=R_{h_{1},k}+\ldots+R_{h_{j_{1}},k}=R_{i_{1},k}+\ldots+R_{i_{j_{2}},k}=H_{2}(R_{k})=0$. The symplectic product of these entries is then 0.
\item  $R_{k}$ is not $P_{k}$ nor $Q_{k}$. As the points are collinear, we have $Q_{k}=P_{k}+R_{k}$. 
Then $H_{1}(Q_{k})=Q_{h_{1},k}+\ldots+Q_{h_{j_{1}},k}=H_{1}(P_{k}+R_{k})=(P_{h_{1},k}+\ldots+P_{h_{j_{1}},k})+(R_{h_{1},k}+\ldots+R_{h_{j_{1}},k})=P_{h_{1},k}+\ldots+P_{h_{j_{1}},k}=H_{1}(P_{k})$ and $H_{2}(Q_{k})=Q_{i_{1},k}+\ldots+Q_{i_{j_{2}},k}=H_{2}(P_{k}+R_{k})=(P_{i_{1},k}+\ldots+P_{i_{j_{2}},k})+(R_{i_{1},k}+\ldots+R_{i_{j_{2}},k})=P_{i_{1},k}+\ldots+P_{i_{j_{2}},k}=H_{2}(P_{k})$, and the $k$-th entries of $\mathit{x}$ and $\mathit{y}$ are $(\alpha, \alpha)$ and $(\beta, \beta)$ and their symplectic product is 0, for all $\alpha$ and $\beta$ in $\mathbb{Z}_{2}$.
\end{itemize}
If $L_{k}$ is such that $S \cap L_{k} = \emptyset$, then one point of the line has to belong to 
$H_{1}$ and another (different from the previous one) has to belong to $H_{2}$. 
The third point of $L_{k}$ belongs to none of the hyperplanes.\\
We can have only one of the following situations.
\begin{itemize}
\item The third point of $L_{k}$ is not $P_{k}$ nor $Q_{k}$. If $H_{1}(P_{k})=1$, i.e. $P_{k}$ 
does not belong to $H_{1}$,  $P_{k}$ has to belong to $H_{2}$, and then $H_{2}(P_{k})=0$; 
moreover $H_{1}(Q_{k})=0$ and $H_{2}(Q_{k})=1$. Instead, if $H_{1}(P_{k})=0$ then $H_{2}(P_{k})=1$, 
$H_{1}(Q_{k})=1$ and $H_{2}(Q_{k})=0$. Briefly, the $k$-th entry of $\mathit{x}$ is 
$(H_{1}(P_{k}),H_{1}(Q_{k}))$, i.e. $(1,0)$ or $(0,1)$, and the  $k$-th entry of $\mathit{y}$ 
is respectively $(0,1)$ or $(1,0)$. The symplectic product of these entries is clearly 1.
\item The third point of $L_{k}$ is $P_{k}$: then $H_{1}(P_{k})=H_{2}(P_{k})=1$, and $Q_{k}$ 
belongs to only one of the hyperplanes, i.e.  $H_{1}(Q_{k})=1$ and $H_{2}(Q_{k})=0$ or $H_{1}(Q_{k})=0$ and $H_{2}(Q_{k})=1$. The $k$-th entries of $\mathit{x}$ and $\mathit{y}$ are $(1, 0)$ and $(1, 1)$ 
or $(1, 1)$ and $(1, 0)$. The symplectic product of these entries is then 1.
\item The third point of $L_{k}$ is $Q_{k}$. We can do the same considerations of the previous point 
and the $k$-th entries of $\mathit{x}$ and $\mathit{y}$ are $(0, 1)$ and $(1, 1)$ or $(1, 1)$ and 
$(0, 1)$. The symplectic product of these entries is then 1.
\end{itemize}
From the above considerations the line $L_{k}$ does not meet the secundum $S$ 
if and only if the symplectic product of the $k$-th coordinate section is 0. 
To calculate the symplectic product of 
two codewords we have to consider only the lines skew to $S$.\\
Now we only have to note that if  $\mathcal{C}$ is a pure quantum code, then all the codewords 
have to be orthogonal each other according to the symplectic product and then, for each 
secundum $S$, the number of lines skew to $S$ must be even. Viceversa if the number 
of lines skew to any secundum $S=H_{1} \cap H_{2}$ is even, then the symplectic product 
between the codewords corresponding to  $H_{1}$ and $H_{2}$ is equal to 0 and the set of 
codewords is a quantum code.
\endproof
According to Definition \ref{parametri} a quantum code is required to be linear
only over $GF(2)$. We can impose the additional condition that the code is linear over
$GF(4)$ as well, i.e. it is closed under multiplication by $\omega$
(where $\omega$ is such that $\omega^{2}+\omega+1=0$). We can
replace each pair $P_{i,k}Q_{i,k}$ in the generator matrix by a corresponding
element of $GF(4)=GF(2)^{2}$. Moreover we can suppose, by
$GF(4)$-linearity, that if  $v_{1},\ldots,v_{m}$ is a $GF(4)$-base
of the code, then $v_{1},\omega v_{1}\ldots,v_{m},\omega v_{m}$ is a
$GF(2)$-base of the same code. Then the generator matrix over
$GF(4)$ of the code is:
\begin{displaymath}
\overline{G}=
\left ( \begin{array}{cccc}
W_{1,1} & W_{1,2} & \ldots&W_{1,n}\\
W_{2,1} & W_{2,2} & \ldots &W_{2,n}\\
\vdots& \vdots & &\vdots\\
W_{m,1} & W_{m,2} & \ldots &W_{m,n}\\
\end{array} \right)
\end{displaymath}
with $W_{i,j} \in \mathbb{F}_{4}\quad \forall i,j$. A $[[n,k,d]]$-quantum code linear over $GF(4)$ can be 
described by a generator matrix of dimensions $\frac{n-k}{2} \times n$.\\
Let $H$ be an hyperplane of $PG(m-1,4)$ of equation:
\begin{displaymath}
H: \alpha_{1}z_{1}+\ldots+\alpha_{m}z_{m}=0 \textrm{ with } \alpha_{i}=a_{i}+\omega b_{i} 
\quad a_{i}, b_{i} \in  \mathbb{F}_{2}\quad  \forall i=1,\ldots,m.
\end{displaymath}
We know that there exits a canonical isomorphism $\Phi : GF(4) \to GF(2)^{2}$ and let $z_{i} \in GF(4)$ be $x_{i}+\omega y_{i}$. Then
we can associate two different hyperplanes of $PG(2m-1,2)$ to $H$:
\begin{displaymath}
\alpha_{1}z_{1}+\ldots+\alpha_{m}z_{m}=0 \iff  (a_{1}+\omega b_{1})(x_{1}+\omega y_{1})+
\ldots+(a_{m}+\omega b_{m})(x_{m}+\omega y_{m})=0
\end{displaymath}
\begin{displaymath}
\iff a_{1}x_{1}+\ldots +a_{m}x_{m}+\omega(b_{1}x_{1}+\ldots+b_{m}x_{m}+a_{1}y_{1}+\ldots+a_{m}y_{m})+\omega^{2}(b_{1} y_{1}+\ldots+b_{m}y_{m})=0
\end{displaymath}
\begin{displaymath}
\iff a_{1}x_{1}+\ldots +a_{m}x_{m}+b_{1} y_{1}+\ldots+b_{m}y_{m} +
\end{displaymath}
\begin{displaymath}
\omega (b_{1}x_{1}+\ldots+b_{m}x_{m}+a_{1}y_{1}+\ldots+a_{m}y_{m}+b_{1} y_{1}+\ldots+b_{m}y_{m})=0
\end{displaymath}
\begin{displaymath}
\iff a_{1}x_{1}+\ldots +a_{m}x_{m}+b_{1} y_{1}+\ldots+b_{m}y_{m}=0 \quad \wedge
\end{displaymath}
\begin{displaymath}
b_{1}x_{1}+\ldots+b_{m}x_{m}+a_{1}y_{1}+\ldots+a_{m}y_{m}+b_{1} y_{1}+\ldots+b_{m}y_{m}=0,
\end{displaymath}
since $\omega^{2}=\omega+1$. Then we can associate to $H$ the following secundum
\begin{equation*}
S: \left \{
\begin{array}{ccc}
a_{1}x_{1}+\ldots +a_{m}x_{m}+b_{1} y_{1}+\ldots+b_{m}y_{m}&=&0\\
b_{1}x_{1}+\ldots+b_{m}x_{m}+(a_{1}+b_{1})y_{1}+\ldots+(a_{m}+b_{m})y_{m}&=&0\\
\end{array}
\right.
\end{equation*}
Clearly not each secundum in $PG(2m-1,2)$ corresponds to an hyperplane of $PG(m-1,4)$: a secundum
\begin{equation*}
S': \left \{
\begin{array}{ccc}
a_{1}x_{1}+\ldots +a_{m}x_{m}+b_{1} y_{1}+\ldots+b_{m}y_{m}&=&0\\
a'_{1}x_{1}+\ldots+a'_{m}x_{m}+b'_{1}y_{1}+\ldots+b'_{m}y_{m}&=&0\\
\end{array}
\right.
\end{equation*}
is a $PG(m-1,4)$-hyperplane $\iff (b_{i}=a'_{i}) \wedge (a_{i}+b_{i}=b'_{i}) \quad \forall i=1,\ldots,m$.\\
The following theorem gives a geometrical description of pure linear quantum codes (see \cite{tesi}).
\begin{theorem}\label{lineare}
The following are equivalent:
\begin{enumerate}
\item A pure quantum $[[n,k,d]]$-code which is linear over $\mathbb{F}_{4}$.
\item A set of $n$ points in $PG(\frac{n-k}{2}-1,4)$ of strength $t=d-1$, such that the intersection 
size with any hyperplane has the same parity as $n$.
\item An $[n,k]_{4}$ linear code of strength $t=d-1$, all of whose weights are even.
\item An $[n,k]_{4}$ linear code of strength $t=d-1$ which is self-orthogonal with respect to the 
Hermitian form.
\end{enumerate}
\end{theorem}
\proof

\indent $1 \Rightarrow 2$. We can utilize Theorem \ref{th:caratt} and the above observations.\\

\indent $2 \Rightarrow 1$. Let $S$ be a secundum of $PG(2m-1,2)$. If it is a 
$GF(4)$-hyperplane we have to prove nothing. So we consider a secundum $S$ which is not an hyperplane of  $PG(m-1,4)$ and
\begin{displaymath}
\omega S=\{ \omega P\in PG(2m-1,2) \textrm{ } | \textrm{ } P \in S\}=
\end{displaymath}
\begin{displaymath}
\{ (y_{1},x_{1}+y_{1},\ldots, y_{m},x_{m}+y_{m}) \in PG(2m-1,2)\textrm{ } | \textrm{ }(x_{1},y_{1},\ldots,x_{m},y_{m}) \in S\}.
\end{displaymath}
$K=S\cap \omega S$  is the greater $GF(4)$-subspace contained in
$S$. In fact $K$ is clearly a subspace of $PG(2m-1,2)$, it is closed
under multiplication by $\omega$ and then is a $GF(4)$-subspace.
Finally every other $GF(4)$-subspace contained in $S$ is closed
under multiplication by $\omega$ and then contained in $\omega S$.
Let $S$ be described by these equations:
\begin{equation*}
S: \left \{
\begin{array}{ccc}
a_{1}x_{1}+\ldots +a_{m}x_{m}+b_{1} y_{1}+\ldots+b_{m}y_{m}&=&0\\
a'_{1}x_{1}+\ldots+a'_{m}x_{m}+b'_{1}y_{1}+\ldots+b'_{m}y_{m}&=&0\\
\end{array}
\right.
\end{equation*}
then $\omega S$ is described by the equations:
\begin{equation*}
\omega S : \left \{
\begin{array}{ccc}
(a_{1}+b_{1})x_{1}+\ldots +(a_{m}+b_{m})x_{m}+a_{1}y_{1}+\ldots +a_{m}y_{m}&=&0\\
(a'_{1}+b'_{1})x_{1}+\ldots +(a'_{m}+b'_{m})x_{m}+a'_{1}y_{1}+\ldots +a'_{m}y_{m}&=&0\\
\end{array}
\right.
\end{equation*}
Then $K$ is described by:
\begin{equation*}
K: \left \{
\begin{array}{ccc}
a_{1}x_{1}+\ldots +a_{m}x_{m}+b_{1} y_{1}+\ldots+b_{m}y_{m}&=&0\\
a'_{1}x_{1}+\ldots+a'_{m}x_{m}+b'_{1}y_{1}+\ldots+b'_{m}y_{m}&=&0\\
(a_{1}+b_{1})x_{1}+\ldots +(a_{m}+b_{m})x_{m}+a_{1}y_{1}+\ldots +a_{m}y_{m}&=&0\\
(a'_{1}+b'_{1})x_{1}+\ldots +(a'_{m}+b'_{m})x_{m}+a'_{1}y_{1}+\ldots +a'_{m}y_{m}&=&0\\
\end{array}
\right.
\end{equation*}
As $S$ is not a $GF(4)$-hyperplane, its four equations are independent to each other and then $K$ 
has binary codimension equal to 4. \\
We know that there exist exactly 5 $GF(4)$-hyperplanes
$H_{1},H_{2},H_{3},H_{4},H_{5} $ which contain $K$ and spread the
remaining points. Let  $n$ be the number of the codepoints, $m$ be
the number of points belonging to $K$ and $a_{i}$ the number of them
contained in $H_{i} \setminus K$. Then we have
\begin{displaymath}
n=m+\sum_{i=1}^{5}a_{i}.
\end{displaymath}
By hypothesis we know that the number of the points not belonging to a $GF(4)$-hyperplane is even. 
In particular for the $H_{i}$ we have that
\begin{displaymath}
n-(m+a_{j})=\sum_{i=1,i\neq j}^{5}a_{i}
\end{displaymath}
is even for each $j=1,\ldots,5$ and then each $a_{i}$ has the same parity. By hypothesis the 
secundum $S$ is not a $GF(4)$-hyperplane, therefore it cannot coincide with any hyperplane $H_{i}$. We can see how the points of $S$ are divided in the hyperplanes $H_{i}$:
\begin{itemize}
\item Let $x_{0} \in S\setminus K$ be a point; then it belongs to some $H_{i_{1}}$ and then 
$K'=(K+x_{0})\cup \{x_{0}\}=\{x+x_{0} \textrm{ }|\textrm{ } x \in K\}\cup \{x_{0}\} \subset H_{i_{1}}$ 
by linearity. In this way we have obtained $|K|+|K|+1=2|K|+1$ points.
\item  Let $y_{0}$ be a point of $S\setminus K'$: then $y_{0}$ is not in $H_{i_{1}}$ 
since we would have $S=H_{i_{1}}$ that is absurd. Then $y_{0} \in H_{i_{2}}$ with $i_{1}\neq i_{2}$ 
and $K''=(K+y_{0})\cup \{y_{0}\} \subset H_{i_{2}}$ by linearity. We have now $3|K|+2$ points.
\item We consider $K'''=K'+y_{0}$: it cannot be contained in $H_{i_{2}}$, because we would have 
$x_{0} \in H_{i_{2}}$, but $x_{0} \in H_{i_{1}}$. Then we have $K''' \subset H_{i_{3}}$, with 
$i_{3}\neq i_{1},i_{2}$. We have obtained $4|K|+3$ points.
\item We have also that  $|S|=2^{m-2}-1$, and $|K|=2^{m-4}-1$. 
Then  $4|K|+3=4(2^{m-4}-1)+3=2^{m-2}-4+3=2^{m-2}-1=|S|$ and $S=K\cup K'\cup K'' \cup K'''$.
\end{itemize}
It is clear that $S$ is contained in three different hyperplanes
$H_{i}$. All the quaternary points belonging to these particular
$GF(4)$-hyperplanes correspond to lines which do not intersect the
$GF(2)$-secundum $S_{i}$ corresponding to $H_{i}$. As
each set $K \cup K'$, $K \cup K''$ e $K \cup K'''$ is a subspace of
$H_{i_{1}}$, $H_{i_{2}}$, $H_{i_{3}}$ of codimension 3, a line
contained in $S_{i_{1}}$, $S_{i_{2}}$ or  $S_{i_{3}}$ must meet
them. Then only the $GF(4)$-points belonging to $H_{j}$, with $j\neq i_{1},i_{2},i_{3}$ correspond to lines which do not intersect $S$ and the number of lines which do not meet $S$ is the sum of two
particular $a_{i}$, and it is even.\\

\indent $ 2 \iff 3$. We consider the correspondence between a codeword and an hyperplane of $PG(m-1,4)$
\begin{displaymath}
\mathit{x}=\alpha_{1}A_{i_{1}}+\ldots+\alpha_{i}A_{i_{j}} \leftrightsquigarrow  H:\alpha_{1}x_{i_{1}}+
\ldots+\alpha_{i}x_{i_{j}}=0.
\end{displaymath}
where $\alpha_{i} \in \mathbb{F}_{4}$. If a point belongs to $H$
then the corresponding entry in the codeword is equal to 0 and
viceversa: the entries not equal to 0 correspond to points not
belonging to the hyperplane. Then if every hyperplane contains a
number of points with the same parity of $n$, the remaining even
points correspond to entries not equal to 0 in the codeword.
Viceversa if a codeword has even weight, i.e. the number of entries
not equal to 0 is even, we have
 the number of the points which do not belong to the hyperplane is even and therefore the number of the 
belonging ones has the same parity with $n$ (total number of points).\\

\indent $3 \iff 4$. A codeword is self-orthogonal respect to the Hermitian product if and only if
\begin{displaymath}
0=(\mathit{x},\mathit{x})\textrm{ mod } 2=
[\sum_{i=1}^{n} x_{i}x_{i}^{2}] \textrm{ mod } 2=[\sum_{i=1}^{n}x_{i}^{3}]
\textrm{ mod } 2=wt(\mathit{x}) \textrm{ mod } 2.
\end{displaymath}
Then $\sum_{i=1}^{n}x_{i}^{3}$ has the same parity of
$w(\mathit{x});$ therefore the weights are even iff all the
codewords are self-orthogonal.
\endproof
In this work we have looked for small complete quantum caps in $PG(4,4)$, which correspond to
 $[[n,n-10,3]]$-codes (see definition \ref{parametri}), using an exhaustive search algorithm.
 Since the computational instruments are the same, we have also looked for the minimum size of
  complete caps in $PG(4,4)$ (see \cite{tesi}). \\

\section{The spectrum of quantum caps in $PG(4,4)$ }\label{spectrum}
In this work we have looked for complete quantum caps in $PG(4,4)$, which correspond to
 $[[n,n-10,4]]$-codes (see Definition \ref{parametri}), using an exhaustive search algorithm helped by theoretical results illustrated in the previous section.
 Since the computational instruments are the same, we have also looked for the minimum size of
  complete caps in $PG(4,4)$ (see \cite{tesi}). \\
\subsection{The searching algorithm }\label{alg}
We start from caps, complete and incomplete, in $PG(3,4)$ where the classification is known, and we try to extend every starting cap joining new points in $PG(4,4)$. The searching algorithm, in C language, organizes the caps in a tree and the extension process ends when the obtained caps are complete. Some considerations about equivalence of caps allow us not to consider, during the process, the caps that will produce caps already found or equivalent to one of these. The algorithm is described in detail in \cite{tesi}.\\

\subsection{Results}\label{risultati}
First of all we have determined, up to equivalence, all the quantum caps in $PG(4,4)$ of sizes $\leq 12$, finding only two examples of 10-incomplete quantum caps and five examples of $12$-incomplete quantum caps. They correspond to $[[10,0,4]]$ and $[[12,2,4]]$-quantum codes.
We already know that there exist quantum caps in $PG(4,4)$ of sizes $[10,12-36,38,40,41]$ (\cite{AMS}, \cite{40Cap}, \cite{41Cap} and \cite{Tonchev08}). \\
Then we have proven by a direct backtracking algorithm that quantum caps of size $11$ do not exist. Successfully we have established the non existence of quantum caps of sizes $37$ and $39$.\\
According to Theorem \ref{lineare} we can consider starting caps in $PG(3,4)$ of odd size only.\\
In particular we consider in our search only caps of sizes $13$, $15$ and $17$ in $PG(3,4)$, since the following theorem and the non existence of particular linear codes.
\begin{theorem}\label{multiset}
The following are equivalent:
\begin{enumerate}
\item An $[n,k,d']_{q}$-code with $d' \geq d$.
\item A multiset $\mathcal{M}$ of points of the projective space $PG(k-1,q)$, 
which has cardinality $n$ and satisfying the following: for every hyperplane $H \subset PG(k-1,q)$ 
there are at least $d$ points of $\mathcal{M}$ outside $H$ (in the multiset sense).
\end{enumerate}
\end{theorem}
More precisely, we know that linear codes with $n =37,39$ $k=5$ and $d > n-12$ do not exist
(see \cite{codetable}) and so there exists an hyperplane which contains at least 12 points of the caps.\\
Then we consider only the examples of non equivalent caps in $PG(3,4)$ contained in the following table:
\begin{table}[h]
\caption{Number and type of non equivalent caps $\mathcal{K} \subset PG(3,4)$, with $|\mathcal{K}|=13,15,17$}
\begin{center}
\begin{tabular}{|c|c|c|}
\hline
$|\mathcal{K}|$&\# COMPLETE&\# INCOMPLETE\\
&CAPS&CAPS\\
\hline
13&1&3\\
15&0&1\\
17&1&0\\
\hline
\end{tabular}
\end{center}
\end{table}

We finish our search, finding no examples of quantum caps in $PG(4,4)$ of sizes $37$ and $39$. According \cite{AMS}, \cite{41Cap}, \cite{40Cap} and \cite{Tonchev08} we have proved the following:
\begin{theorem}
If $\mathcal{K} \subset PG(4,4)$ is a quantum cap, then $10\leq |\mathcal{K}| \leq 41$, with $|\mathcal{K}| \neq 11,37,39$.
\end{theorem}

\subsection{List of found caps}
We list all the non equivalent quantum caps of sizes $\leq 12$ and some examples of complete quantum caps of sizes $20, 29, 30, 32, 33, 34, 36, 38$. These examples are not equivalent to those constructed in \cite{Tonchev08}, since the last are subsets of the $41$-complete quantum cap. Let $\mathbb{F}_{4}=\{0,1,\omega,\omega^{2}\}$. In the following list we will write for brevity $2=\omega$ and $3=\omega^{2}$.
\subsubsection{10-incomplete quantum caps}

\begin{minipage}[t]{8,25 cm}
\begin{center}
CAP 1
\end{center}
\scriptsize
\begin{center}
\begin{tabular}{ccccccccccc}
 1& 0& 0& 0& 1& 1& 0& 0& 1& 0\\
 0& 1& 0& 0& 1& 1& 0& 0& 0& 1\\
 0& 0& 1& 0& 1& 1& 0& 1& 0& 0\\
 0& 0& 0& 1& 3& 1& 0& 2& 2& 2\\
 0& 0& 0& 0& 2& 1& 1& 3& 3& 3\\
 \end{tabular}
 \end{center}
\normalsize
\begin{center}
Its weight distribution is:
 {\scriptsize $[<4, 30>,< 6,300>,<8,585>,<10,108>] $}
 \end{center}
 \end{minipage}
 \begin{minipage}[t]{8,25 cm}
  \begin{center}
 CAP 2
\end{center}
\scriptsize
  \begin{center}
\begin{tabular}{cccccccccc}
 1& 0& 1& 0& 0& 0& 1& 0& 0& 1\\
 0& 1& 3& 0& 0& 0& 1& 0& 0& 2\\
 0& 0& 2& 1& 0& 0& 1& 1& 1& 3\\
 0& 0& 2& 0& 1& 0& 1& 2& 3& 3\\
 0& 0& 2& 0& 0& 1& 1& 3& 2& 3\\
\end{tabular}
\end{center}
\normalsize
 \begin{center}
Its weight distribution is:
{\scriptsize $[<4,30>,<6,300>,<8,585>,<10,108>]$}
\end{center}
\end{minipage}

\subsubsection{12-incomplete quantum caps}
They correspond to $[[12,2,4]]$-quantum codes.\\

\begin{minipage}[t]{8,25 cm}
\begin{center}
CAP 1
\end{center}
\scriptsize
\begin{center}
\begin{tabular}{cccccccccccc}
 1& 1& 0& 0& 0& 0& 0& 1& 0& 1& 1& 1\\
 0& 2& 1& 0& 1& 0& 0& 1& 0& 0& 2& 1\\
 0& 0& 0& 1& 0& 0& 0& 1& 1& 2& 1& 2\\
 0& 2& 0& 0& 2& 1& 0& 1& 1& 3& 1& 1\\
 0& 2& 0& 0& 2& 0& 1& 1& 2& 0& 3& 0\\
\end{tabular}
 \end{center}
\normalsize
\begin{center}
Its weight distribution is:
 {\scriptsize $[<6,84>,<8,405>,<10,468>,<12,66>] $}
 \end{center}
 \end{minipage}
 \begin{minipage}[t]{8,25 cm}
  \begin{center}
 CAP 2
\end{center}
\scriptsize
  \begin{center}
\begin{tabular}{cccccccccccc}
 1& 0& 0& 0& 0& 1& 0& 1& 0& 1& 0& 0\\
 0& 1& 0& 0& 0& 3& 0& 1& 1& 2& 1& 0\\
 0& 0& 1& 1& 0& 1& 0& 1& 0& 0& 3& 1\\
 0& 0& 1& 0& 1& 3& 0& 1& 2& 2& 3& 2\\
 0& 0& 2& 0& 0& 3& 1& 1& 2& 2& 3& 1\\
\end{tabular}
\end{center}
\normalsize
 \begin{center}
Its weight distribution is:
{\scriptsize $[<4,6>,<6,60>,<8,441>,<10,444>,<12,72>]$}
\end{center}
\end{minipage}

\vspace*{1 cm}
\begin{minipage}[t]{8,25 cm}
\begin{center}
CAP 3
\end{center}
\scriptsize
\begin{center}
\begin{tabular}{cccccccccccc}
 0& 1& 1& 0& 0& 0& 0& 0& 0& 1& 1& 0\\
 0& 0& 1& 1& 1& 0& 0& 0& 1& 1& 0& 1\\
 1& 0& 0& 0& 3& 1& 0& 0& 2& 1& 1& 0\\
 1& 0& 0& 0& 3& 0& 1& 0& 0& 1& 1& 2\\
 2& 0& 2& 0& 3& 0& 0& 1& 2& 1& 3& 2\\
\end{tabular}
 \end{center}
\normalsize
\begin{center}
Its weight distribution is:
 {\scriptsize $[<4,6>,<6,60>,<8,441>,<10,444>,<12,72>]$}
 \end{center}
 \end{minipage}
 \begin{minipage}[t]{8,25 cm}
  \begin{center}
 CAP 4
\end{center}
\scriptsize
  \begin{center}
\begin{tabular}{cccccccccccc}
0& 1& 0& 0& 0& 0& 0& 0& 1& 1& 0& 1\\
 0& 0& 1& 1& 0& 0& 1& 0& 1& 0& 1& 1\\
 1& 0& 0& 0& 1& 0& 2& 0& 1& 1& 3& 0\\
 1& 0& 0& 2& 0& 1& 0& 0& 1& 1& 3& 0\\
 2& 0& 0& 1& 0& 0& 1& 1& 1& 0& 0& 1\\
\end{tabular}
\end{center}
\normalsize
 \begin{center}
Its weight distribution is:
{\scriptsize $[<4,9>,<6,48>,<8,459>,<10,432>,<12,75>]$}
\end{center}
\end{minipage}

\vspace*{1 cm}
\begin{center}
CAP 5
\end{center}
\scriptsize
\begin{center}
\begin{tabular}{cccccccccccc}
 1& 0& 0& 0& 0& 1& 0& 1& 0& 0& 1& 0\\
 0& 1& 0& 0& 0& 1& 0& 1& 0& 0& 0& 1\\
 0& 0& 1& 0& 0& 1& 1& 3& 1& 1& 2& 2\\
 0& 0& 0& 1& 0& 1& 3& 3& 1& 2& 2& 2\\
 0& 0& 0& 0& 1& 1& 3& 3& 2& 1& 2& 2\\
\end{tabular}
\end{center}
\normalsize
\begin{center}
Its weight distribution is:\\
{\scriptsize $[<4,18>,<6,12>,<8,513>,<10,396>,<12,84>]$}
\end{center}

\subsubsection{The 20-complete quantum cap}\label{20quantumcap}
We have also found other quantum caps by the searching algorithm described in \S  \ref{alg}.
Starting from caps in $PG(3,4)$ and utilizing the procedure
described in the previous section  we have determined a 20-quantum
cap; it has been obtained starting from a 12-complete cap in
$PG(3,4)$. The coordinates of this cap are the
following:
\scriptsize
\begin{center}
\begin{tabular}{cccccccccccccccccccc}
0&   1&   0&   0&   0&   0&   0&   0&   1&   1&   0&   1&   0&   0&   0&   1& 1&   1&   0&   1\\
0&   0&   1&   1&   0&   1&   0&   0&   2&   1&   0&   1&   1&   0&   1&   0& 2&   3&   1&   3\\
1&   0&   0&   0&   1&   3&   0&   0&   2&   1&   1&   2&   1&   1&   1&   3& 1&   0&   3&   3\\
1&   0&   0&   2&   0&   1&   1&   0&   3&   3&   2&   2&   0&   3&   1&   1& 2&   1&   2&   0\\
2&   0&   0&   1&   0&   2&   0&   1&   0&   3&   1&   1&   2&   3&   1&   2& 2&   1&   0&   3\\
\end{tabular}
\end{center}
\normalsize
\begin{center}
Its weight distribution is: \\
{ \scriptsize$ [<0, 1>, <8, 3>, <12, 117>, <14, 432>, <16, 312>, <18, 144>, <20, 15> ]$}
\end{center}

\noindent As all the weights are even this cap is quantic by  Theorem \ref{lineare}.
This cap generates a $[[20,10,4]]$-quantum code. The size of its  \emph{stabilizer} is 48 and it is generated by the following projectivities:
\footnotesize
\begin{equation*}
G_{1} =\left (
\begin{array}{ccccc}

1&0&0&0&0\\

0&0&0&\overline{\omega}&0\\

0&\omega&0&\overline{\omega}&\omega\\

0&\overline{\omega}&\omega&1&0\\

0&1&\overline{\omega}&1&\omega\\

\end{array}
\right )
G_{2}= \left (
\begin{array}{ccccc}
1&0&\overline{\omega}&1&\omega\\

0&1&0&0&0\\

0&0&1&0&0\\

0&0&0&1&0\\

0&0&0&0&1\\

\end{array}
\right )
G_{3}= \left (
\begin{array}{ccccc}
1&\omega&\omega&\overline{\omega}&0\\

0&1&0&0&0\\

0&0&1&0&0\\

0&0&0&1&0\\

0&0&0&0&1\\
\end{array}
\right ).
\end{equation*}
\normalsize
\subsubsection{29-complete quantum cap}
This 29-complete cap is obtained from a 17-complete cap in $PG(3,4)$. It corresponds to an $[[29,19,4]]$-quantum code.
\scriptsize
\begin{center}
\begin{tabular}{ccccccccccccccccccccccccccccc}
1& 0& 0& 0& 0& 1& 0& 1& 1& 1& 1& 0& 0& 0& 0& 0& 0& 0& 1& 1& 0& 1& 0& 0& 0& 0& 
1& 1& 1\\
 0& 1& 0& 0& 0& 0& 1& 0& 3& 0& 0& 1& 0& 1& 1& 1& 0& 1& 2& 0& 1& 1& 1& 1& 1& 1& 
2& 1& 3\\
 0& 0& 1& 0& 0& 0& 2& 3& 0& 1& 1& 1& 1& 0& 2& 3& 1& 1& 2& 0& 3& 3& 0& 3& 1& 2& 
2& 1& 1\\
 0& 0& 0& 1& 0& 1& 0& 1& 1& 0& 1& 2& 1& 2& 3& 3& 3& 1& 2& 2& 2& 1& 3& 1& 0& 1& 
3& 1& 3\\
 0& 0& 0& 0& 1& 3& 1& 2& 1& 2& 0& 0& 2& 1& 2& 0& 3& 1& 0& 2& 3& 0& 3& 2& 2& 3& 
3& 2& 1\\
\end{tabular}
\end{center}
\normalsize
\begin{center}
Its weight distribution is:\\
 { \scriptsize$ [<12  , 3 >, < 18  , 42 > ,<  20  , 360 >, < 22  , 420> , < 24  , 81> ,<  26  , 90>,<  28  , 27> ]$}\\
 \end{center}

\noindent The following 29-complete cap is obtained from a 13-incomplete cap in $PG(3,4)$. It corresponds to an $[[29,19,4]]$-quantum code.
\scriptsize
\begin{center}
\begin{tabular}{ccccccccccccccccccccccccccccc}
1& 1& 1& 0& 0& 0& 0& 0& 1& 0& 1& 1& 1& 1& 1& 0& 0& 0& 0& 1& 1& 1& 0& 1& 0& 1& 
1& 1& 0\\
 2& 0& 0& 1& 0& 0& 0& 1& 1& 0& 2& 1& 2& 1& 0& 1& 0& 1& 1& 3& 3& 1& 0& 3& 1& 3& 
3& 1& 1\\
 1& 0& 0& 0& 1& 0& 0& 1& 0& 1& 2& 2& 1& 2& 3& 1& 1& 0& 0& 2& 1& 3& 1& 3& 1& 0& 
2& 0& 0\\
 2& 0& 2& 0& 0& 1& 0& 0& 0& 2& 3& 1& 3& 0& 0& 2& 1& 1& 2& 3& 0& 3& 3& 3& 1& 1& 
2& 1& 3\\
 0& 0& 1& 0& 0& 0& 1& 3& 1& 1& 2& 1& 1& 0& 1& 2& 2& 2& 1& 3& 3& 1& 3& 2& 1& 2& 
2& 0& 3\\\end{tabular}
\end{center}
\normalsize
\begin{center}
Its weight distribution is:\\
{\scriptsize
 $[  <16  , 6 > ,<  18  , 57 > , < 20  , 348 >, < 22  , 366 > , < 24  , 159 >, < 26  , 57 > , < 28  ,30 >]$}
 \end{center}

\subsubsection{30-complete quantum cap}
This 30-complete cap is obtained from a 16-complete cap in $PG(3,4)$. It corresponds to an $[[30,20,4]]$-quantum code.
\scriptsize
\begin{center}
\begin{tabular}{ccccccccccccccc}
 1& 1& 0& 0& 1& 0& 0& 0& 1& 1& 1& 0& 0& 1& 0\\
 0& 0& 1& 0& 1& 0& 0& 1& 3& 1& 0& 1& 0& 2& 1\\
 0& 0& 0& 1& 1& 0& 0& 2& 0& 3& 1& 1& 1& 1& 0\\
 0& 2& 0& 0& 1& 1& 0& 0& 2& 1& 2& 2& 1& 3& 2\\
 0& 1& 0& 0& 3& 0& 1& 1& 0& 1& 2& 0& 2& 2& 1\\
\\
 0& 1& 0& 0& 1& 0& 1& 0& 1& 0& 0& 1& 0& 1& 1\\
 1& 1& 0& 1& 2& 1& 0& 1& 1& 1& 1& 2& 1& 0& 1\\
 2& 0& 1& 1& 2& 3& 1& 0& 2& 3& 1& 1& 2& 0& 2\\
 3& 2& 3& 1& 0& 2& 0& 3& 2& 1& 0& 1& 1& 1& 1\\
 2& 2& 3& 1& 2& 3& 3& 3& 1& 2& 2& 1& 3& 2& 2\\
\end{tabular}
\end{center}
\normalsize
\begin{center}
Its weight distribution is:\\
 {\scriptsize $[ <14 , 3 >,  <20  ,258 > , < 22  , 438 > , < 24  , 165 > , <26  , 108 > ,< 28 , 48> ,<  30  , 3 > ]$}
 \end{center}
 
\subsubsection{32-complete quantum cap}
This 32-complete cap is obtained from a 16-complete cap in $PG(3,4)$. It corresponds to an $[[32,22,4]]$-quantum code. The research is not complete.
\scriptsize
\begin{center}
\begin{tabular}{cccccccccccccccc}
 1& 1& 0& 0& 1& 0& 0& 0& 1& 1& 1& 0& 0& 0& 0& 0\\
 3& 0& 1& 0& 1& 0& 0& 1& 2& 2& 2& 1& 0& 1& 1& 0\\
 3& 0& 0& 1& 1& 0& 0& 2& 2& 0& 1& 1& 1& 0& 2& 1\\
 2& 0& 0& 0& 1& 1& 0& 0& 1& 1& 1& 2& 1& 2& 3& 3\\
 0& 0& 0& 0& 3& 0& 1& 1& 1& 3& 0& 0& 2& 1& 2& 3\\
\\
 0& 1& 1& 0& 1& 1& 0& 1& 1& 1& 0& 0& 1& 1& 0& 1\\
 1& 0& 2& 1& 0& 1& 1& 0& 0& 1& 1& 1& 3& 2& 1& 3\\
 1& 0& 3& 3& 1& 0& 0& 3& 1& 2& 1& 3& 2& 0& 2& 1\\
 1& 2& 1& 2& 0& 1& 3& 1& 2& 2& 0& 1& 2& 0& 1& 3\\
 1& 2& 2& 3& 3& 0& 3& 0& 1& 1& 2& 2& 3& 1& 3& 0\\
\end{tabular}
\end{center}
\normalsize
\begin{center}
Its weights distribution is: {\scriptsize$[<16  , 3 >,<  20  , 39 >,  <22  , 312 >,  <24  , 429 >, < 26  , 120>, < 28  , 69>, < 30  , 48>,< 32 ,3 >]$}
\end{center}

\subsubsection{33-complete quantum caps}
We have found these 3 non equivalent quantum caps of size 33, starting from an incomplete cap of size 13 in $PG(3,4)$. The research is not complete. They correspond to $[[33,23,4]]$-quantum codes.\\

\scriptsize
\begin{center}
\begin{tabular}{lccccccccccccccccc}
& 1& 0& 1& 0& 0& 0& 1& 0& 1& 1& 0& 1& 1& 1& 1& 1& 1\\
& 0& 1& 2& 0& 0& 0& 2& 1& 2& 1& 0& 2& 1& 0& 1& 3& 3\\
& 0& 0& 1& 1& 0& 0& 0& 1& 2& 3& 1& 2& 2& 1& 2& 0& 0\\
& 0& 0& 0& 0& 1& 0& 2& 0& 1& 0& 2& 3& 1& 0& 2& 0& 3\\
& 0& 0& 1& 0& 0& 1& 1& 3& 0& 2& 1& 2& 1& 2& 0& 1& 0\\
{\normalsize CAP 1}\\
& 0& 0& 0& 0& 1& 1& 0& 0& 1& 1& 1& 1& 0& 1& 1& 1\\
& 1& 0& 1& 1& 1& 0& 0& 1& 0& 3& 3& 2& 1& 0& 3& 1\\
& 1& 1& 0& 0& 3& 2& 1& 1& 0& 3& 1& 3& 0& 3& 1& 1\\
& 2& 1& 2& 1& 3& 3& 3& 1& 2& 2& 2& 1& 3& 1& 1& 3\\
& 2& 2& 1& 2& 1& 1& 3& 1& 2& 2& 1& 2& 3& 1& 2& 2\\
\end{tabular}
\end{center}
\normalsize
\begin{center}
Its weights distribution is:\\
 {\scriptsize$[<18  , 3 > , < 20  ,6> , < 22  , 204>, < 24  , 435>, < 26  , 219> , < 28  , 84> , < 30 , 54 >, < 32  , 18 > ]$}
\end{center}
\scriptsize
\begin{center}
\begin{tabular}{lccccccccccccccccc}
& 1& 1& 1& 0& 0& 1& 0& 0& 0& 1& 0& 0& 1& 1& 0& 1& 1\\ 
& 0& 2& 1& 1& 0& 2& 0& 0& 1& 2& 0& 0& 0& 2& 1& 1& 1\\
& 0& 1& 3& 0& 1& 1& 0& 0& 0& 1& 1& 1& 0& 2& 1& 2& 2\\
& 0& 0& 2& 0& 0& 3& 1& 0& 3& 1& 2& 1& 2& 0& 1& 3& 1\\
& 0& 2& 0& 0& 0& 3& 0& 1& 3& 1& 1& 2& 3& 0& 1& 3& 1\\
{\normalsize CAP 2}\\
& 1& 1& 0& 1& 1& 1& 1& 0& 0& 1& 1& 1& 0& 0& 1& 1\\
& 3& 2& 1& 1& 1& 0& 3& 1& 1& 3& 2& 0& 0& 1& 2& 0\\
& 0& 2& 0& 1& 3& 0& 0& 1& 1& 3& 1& 2& 1& 0& 2& 2\\
& 2& 3& 1& 1& 3& 1& 0& 2& 0& 3& 2& 1& 3& 2& 2& 0\\
& 1& 2& 2& 3& 1& 1& 3& 2& 3& 0& 0& 0& 3& 1& 3& 1\\
\end{tabular}
\end{center}
\normalsize
\begin{center}
Its weights distribution is:\\
  {\scriptsize$[<16  , 3>,<  20  , 27 >, < 22  , 108> , < 24  , 573>, < 26  , 144> ,<  28  , 105 >,<  30  , 36> ,< 32  , 27 >] $}
  \end{center}

\scriptsize
\begin{center}
\begin{tabular}{lccccccccccccccccc}
&1& 1& 0& 0& 0& 1& 0& 1& 0& 0& 0& 1& 1& 1& 0& 1& 1\\
&0& 1& 1& 0& 0& 3& 0& 3& 1& 0& 0& 0& 0& 2& 1& 1& 1\\
&0& 3& 0& 1& 0& 1& 0& 1& 0& 1& 1& 3& 0& 2& 1& 2& 2\\
&0& 2& 0& 0& 1& 2& 0& 3& 3& 2& 1& 2& 2& 0& 1& 3& 1\\
&0& 0& 0& 0& 0& 0& 1& 1& 3& 1& 2& 1& 3& 0& 1& 3& 1\\
{\normalsize CAP 3}\\
&1& 0& 1& 1& 1& 1& 1& 0& 0& 1& 1& 1& 0& 0& 1& 1\\
&2& 1& 0& 2& 1& 1& 0& 1& 1& 3& 3& 3& 0& 1& 2& 2\\
&2& 0& 3& 0& 1& 3& 0& 1& 1& 3& 1& 1& 1& 0& 2& 0\\
&3& 1& 0& 0& 1& 3& 1& 2& 0& 3& 1& 0& 3& 2& 2& 1\\
&2& 2& 3& 1& 3& 1& 1& 2& 3& 0& 2& 3& 3& 1& 3& 0\\
\end{tabular}
\end{center}
\normalsize
\begin{center}
Its weights distribution is: \\
 {\scriptsize$[<16  , 3 > , <20  , 18 > , <22  , 144>  ,<24  , 516 > ,<26  , 192>  , <28  , 78 > ,<30  , 48>  ,<32  , 24>]$}
 \end{center}

\subsubsection{34-complete quantum caps}
We have found over 130 non equivalent quantum caps of size 34, starting from cap of size 16 in $PG(3,4)$. The research is not complete. They correspond to $[[34,24,4]]$-quantum codes.

\subsubsection{36-complete quantum caps}
These 36-complete caps are obtained from a 16-complete cap in $PG(3,4)$
They correspond to $[[36,26,4]]$-quantum codes. 
\scriptsize
\begin{center}
\begin{tabular}{lcccccccccccccccccc}
& 1& 1& 1& 0& 0& 0& 0& 0& 0& 0& 0& 0& 1& 1& 1& 0& 0& 1\\
& 0& 2& 1& 1& 0& 1& 0& 0& 1& 1& 1& 0& 2& 1& 0& 1& 1& 2\\
& 0& 1& 3& 0& 1& 3& 0& 0& 3& 0& 1& 1& 3& 1& 1& 2& 1& 3\\
& 0& 0& 2& 0& 0& 2& 1& 0& 1& 3& 2& 1& 3& 0& 0& 3& 1& 1\\
& 0& 2& 0& 0& 0& 3& 0& 1& 2& 3& 0& 2& 1& 1& 3& 2& 1& 2\\
{\normalsize CAP 1}\\
& 1& 1& 0& 0& 1& 1& 1& 1& 1& 1& 1& 1& 1& 1& 1& 0& 0& 0\\
&0& 0& 1& 1& 1& 2& 1& 3& 3& 1& 2& 3& 3& 3& 0& 0& 1& 1\\
&2& 0& 1& 2& 1& 0& 3& 2& 0& 2& 1& 2& 0& 3& 2& 1& 0& 2\\
&3& 1& 0& 1& 1& 2& 0& 2& 2& 2& 1& 0& 3& 0& 1& 3& 2& 0\\
&3& 2& 2& 3& 3& 2& 3& 2& 0& 1& 0& 1& 2& 0& 0& 3& 1& 1\\
\end{tabular}
\end{center}
\normalsize
\begin{center}
Its weight distribution:\\
  {\scriptsize$[<20,6>,<24,138>,<26,492>,<28,234>,<30,48>,<32,69>,<34,36>]$}
  \end{center}

\scriptsize
\begin{center}
\begin{tabular}{lcccccccccccccccccc}
& 1& 0& 0& 1& 0& 0& 1& 0& 1& 1& 1& 0& 0& 1& 0& 1& 1& 0\\
& 0& 1& 0& 1& 0& 0& 1& 1& 3& 3& 3& 1& 0& 0& 1& 0& 3& 1\\
& 0& 0& 1& 1& 0& 0& 3& 2& 0& 0& 2& 1& 1& 2& 0& 3& 3& 2\\
& 0& 0& 0& 1& 1& 0& 0& 0& 2& 3& 0& 2& 1& 1& 2& 3& 0& 3\\
& 0& 0& 0& 3& 0& 1& 3& 1& 0& 2& 1& 0& 2& 0& 1& 2& 0& 2\\
{\normalsize CAP 2}\\
&0& 0& 1& 1& 0& 1& 1& 0& 1& 0& 0& 1& 0& 1& 1& 1& 1& 1\\
&0& 1& 2& 3& 1& 0& 1& 1& 1& 1& 1& 1& 1& 1& 2& 1& 0& 2\\
&1& 1& 0& 2& 3& 1& 0& 0& 2& 1& 3& 3& 2& 1& 1& 1& 2& 3\\
&3& 1& 2& 2& 2& 0& 1& 3& 2& 0& 1& 2& 1& 0& 3& 3& 0& 3\\
&3& 1& 2& 2& 3& 3& 0& 3& 1& 2& 2& 0& 3& 1& 3& 2& 2& 1\\
\end{tabular}
\end{center}
\normalsize
\begin{center}
Its weight distribution is:\\
  {\scriptsize$[<20,6>,<22,6>,<24,120>,<26,510>,<28,222>,<30,66>,<32,51>,<34,42>]$}
  \end{center}

\subsubsection{38-complete quantum cap}
This 36-complete cap is obtained from a 16-complete cap in $PG(3,4)$
It corresponds to $[[38,28,4]]$-quantum code. 

\scriptsize
\begin{center}
\begin{tabular}{ccccccccccccccccccc}
 1& 0& 0& 1& 0& 1& 0& 0& 1& 1& 1& 0& 0& 1& 1& 1& 0& 1& 0\\
 0& 1& 0& 2& 0& 0& 0& 1& 3& 2& 3& 0& 0& 1& 1& 1& 1& 0& 1\\
 0& 0& 1& 1& 0& 1& 0& 0& 2& 3& 0& 1& 1& 1& 2& 0& 1& 0& 0\\
 0& 0& 0& 3& 1& 0& 0& 3& 3& 0& 1& 2& 1& 0& 0& 2& 1& 1& 1\\
 0& 0& 0& 3& 0& 2& 1& 3& 2& 0& 0& 1& 2& 1& 3& 3& 1& 2& 2\\
\\
1& 1& 1& 0& 1& 1& 1& 1& 1& 1& 0& 0& 1& 1& 1& 0& 0& 1& 1\\
0& 1& 0& 1& 2& 3& 0& 1& 1& 3& 1& 1& 2& 3& 2& 0& 1& 2& 3\\
3& 1& 3& 1& 2& 3& 1& 2& 0& 0& 1& 1& 1& 3& 3& 1& 0& 2& 2\\
0& 1& 2& 3& 3& 2& 1& 2& 3& 3& 2& 0& 1& 3& 1& 3& 2& 2& 2\\
3& 3& 0& 0& 1& 2& 1& 0& 0& 3& 2& 3& 0& 0& 3& 3& 1& 3& 1\\
\end{tabular}
\end{center}
\normalsize
\begin{center}
Its weight distribution is:\\  {\scriptsize$[<22,6>,<24,12>,<26,288>,<28,288>,<30,372>,<32,3>,<36,48>,<38,6>]$}
\end{center}

\subsection{Minimum size of complete caps in $PG(4,4)$}\label{OrdineMinimo}
As we have already said, since the computational instruments are the same, we also searched for the minimum size of complete caps in $PG(4,4)$.
For this research the smallest size of the caps in $PG(3,4)$ considered is 8, since Theorem \ref{multiset} and the non existence of particular linear codes.\\
More precisely, we know that the minimum size of complete caps in $PG(4,4)$ is at least 19 
(see \cite{18primo} and \cite{18secondo}) and linear codes with $n \geq 19$, $k=5$ and 
$d \geq n-8$ do not exist (see \cite{codetable}), so there exists an hyperplane which contains 
at least 8 points of the caps. \\
We have searched
exhaustively complete caps of size 19. As the research has not given
results and 20-complete caps exist (see \cite{FainaPambianco}), we have proven the following 
(see \cite{OrdineMinimo}):
\begin{theorem}
The minimum size of complete caps in $PG(4,4)$ is 20.
\end{theorem}
In particular we have found a new example of $20$ complete cap (see \S$\textrm{ }$\ref{20quantumcap}) which is quantic.



\end{document}